\documentstyle[12pt]{article}
\begin{document}{}
\begin{titlepage}
\begin{flushright}
UBCTP--93--024 preprint \today\\[0.25in]
\end{flushright}
\begin{center}
{\Large\bf A Kucha\v{r} Hypertime Formalism
For Cylindrically\\
Symmetric Spacetimes
With Interacting Scalar Fields}
{\\[0.5in] \large  Stephen P\@. Braham}
{\\[0.5in] \em Department of Physics
\\University of British Columbia
\\Vancouver, BC, Canada V6T 1Z1
\\E-mail: braham@physics.ubc.ca\\[0.5in]}
\end{center}
\centerline{\large\bf Abstract}
\vskip 0.25in
{\small
The Kucha\v{r} canonical transformation for vacuum geometrodynamics in
the presence of cylindrical symmetry is applied to a general
non-vacuum case. The resulting constraints are highly non-linear and non-local
in the momenta conjugate to the Kucha\v{r} embedding variables. However,
it is demonstrated that the constraints can be
solved for these momenta and thus the
dynamics of cylindrically symmetric models can be cast in a form
suitable for the construction of a hypertime functional Schr\"odinger
equation.

}
\end{titlepage}

The Kucha\v{r} embedding variable approach, in both its internal \cite{KK3}
and external \cite{DIFIK1,DIFIK2,GCS} forms holds great promise for being
a useful arena for the discussion of many problems in
geometrodynamics, particularly in the quantum theory. The approach,
however, has only been applied to models \cite{KK1,CCO,KT2} in which the
matter field content is trivial and we need to start applying it to more
realistic models if we hope to reap its potential advantages. In particular,
the construction in the specific model used by Kucha\v{r} relied
on the conformal invariance of the Einstein-Rosen wave \cite{KK1,CCO},
whereas the standard spherical models for black-holes manifestly break this
invariance \cite{NOTES,SSFII}.
Here we report on the construction of a classical Hamiltonian internal
hypertime formulation for a large class of $2D$ gravity models, with
interacting fields.
This is the first step in the development of a quantum theory based on a
hypertime functional Schr\"odinger equation \cite{KK3,KK1} for these,
and more general, $2D$ models.

The primary steps in such a formulation, described by Kucha\v{r} \cite{KK3},
start with the standard ADM \cite{ADM} formulation for geometrodynamics,
with the action functional being written in the form
\begin{equation}{}\label{hamac}
S = \int dt \, \int d^3x \left(
             {\pi}^{ij}({\bf x}) {\dot g}_{ij}({\bf x})
             - N ({\bf x}) {\cal H} ({\bf x})
             - N^i ({\bf x}) {\cal H}_i ({\bf x}) \right),
\end{equation}
with the canonical coordinates $g_{ij}$ being the metric on
a spacelike hypersurface (with coordinate ${\bf x}$ and label $t$),
$\pi_{ij}$ being the corresponding momenta and overdot represents
partial differentiation with respect to time. The variation of the
embedding of this surface into the spacetime is represented by
the variation of the lapse $N$ and shift $N^i$, giving rise to
the standard superhamiltonian and supermomentum initial value
constraints
\begin{equation}{}\label{coneqs}
{\cal H} ({\bf x}) = 0,\ {\cal H}_i ({\bf x})=0
\end{equation}
respectively. It is the vanishing of the constraints that forces
the vanishing of the Hamiltonian for geometrodynamics, and thus
leads us to the loss of time evolution. The first step is to
make a canonical transformation and divide the variables into
three classes \cite{KK3}
\begin{equation}{}\label{transform}
g_{ij},\pi^{ij} \rightarrow T^\mu, \Pi_\mu, Y_\mu \equiv \{ g^A, \pi_A \},
\end{equation}
where $\mu$ ranges from one to four,
$A \in \{1,2\}$, $g^A$ and $\pi_A$ represent
the `true' dynamical variables of the theory, $T^\mu$ represents
internal embedding coordinates, describing the location of the
hypersurface in spacetime and $\Pi_\mu$ are the corresponding momenta
(i.e.\ the energy-momentum densities). The superhamiltonian
and supermomenta will then generally become non-local functionals
of the new variables. The idea is to then solve
the constraints given by Equation~(\ref{coneqs}) for the embedding
momenta in terms of the other variables
\begin{equation}{}\label{solnPi}
\Pi_\mu ({\bf x}) = -P_\mu ({\bf x}) [T^\nu, Y_\rho],
\end{equation}
where we explicitly denote the possible functional dependence. This
solution, however, is often not unique.
Kucha\v{r} then shows \cite{KK3} that the dynamical variables obey the
following hypertime functional Hamilton equation:
\begin{equation}{}\label{HHam}
\frac{\delta Y_\mu}{\delta T^\nu} = [ Y_\mu, P_\nu ]_P,
\end{equation}
where $[,]_P$ is the Poisson bracket for only the $Y^\mu$
variables. This equation is solved by
prescribing some `internal path', which is simply a specification
of $T^\mu ({\bf x},t)$, $t_i \leq t \leq t_f$ with $T^\mu ({\bf x},t_i)$
representing the position of the
initial hypersurface and $T^\mu ({\bf x},t_f)$ the position of the final
hypersurface. We then have
\begin{equation}{}\label{THam}
\frac{\partial Y_\mu ({\bf x},t)}{\partial t}
= \int d^3x'\, [ Y_\mu ({\bf x},t), P_\nu ({\bf x'},t) ]_P\,
  \dot{T^\nu} ({\bf x'},t),
\end{equation}
with initial data of the form $Y_\mu ({\bf x},t_0) =
Y_\mu (T^\nu ({\bf x},t_0))$.
This equation is then automatically invariant to the choice of internal
path $T^\nu (t)$ between the initial and final configurations,
as discussed by Kucha\v{r} \cite{KK3}, and thus we can view
Equation~(\ref{HHam}) as a valid hypertime evolution equation determining
$Y_\mu$ as a functional of $T^\nu$.

Our model will be based on the cylindrically symmetric spacetimes
first studied by Einstein and Rosen \cite{ER}. Kucha\v{r} \cite{KK1}
has studied the vacuum case for these models, and constructed a canonical
tranformation that allowed for a local solution of Equation~(\ref{solnPi}).
Torre \cite{CCO} has demonstrated how one may then construct the quantum
observables for the model. We start with the cylindrically symmetric
metric,
\begin{equation}{}\label{ERmet}
ds^2 = \exp (2\gamma-4f_{(0)}) (-\alpha^2 dt^2+2\beta dr\,dt + dr^2)
       + \phi^2 \exp (-4f_{(0)}) d\theta^2 + \exp (4f_{(0)}) dz^2,
\end{equation}
where the metric functions $\gamma$, $f_{(0)}$, $\alpha$, $\beta$ and $\phi$
are functions only of $r$ and $t$, we have $0 \leq \theta <0$, and
the boundary conditions are those described by Kucha\v{r} \cite{KK1}.
We then couple a set of $N$ scalar fields $f_{(i)}$ to this model
with an arbitrary self-interaction, resulting in an action of
the form
\begin{equation}{}\label{act}
S=-\int dt\, \int dr\, \phi \sqrt{-g} \left( \frac{1}{8}R + \sum_{i=0}^{N}
\nabla^a f_{(i)} \nabla_a f_{(i)} + V(f_{(i)}) \right),
\end{equation}
where and $g$, $R$ and $\nabla^a$
represent the metric determinant, curvature and covariant derivative
respectively on the two-dimensional space with metric
\begin{equation}{}\label{2dm}
{}^{(2)}ds^2 = \exp (2\gamma) (-\alpha^2 dt^2+2\beta dr\,dt + dr^2),
\end{equation}
and $V(f_{(i)})$ represents the interaction, including the factor coming
from the Einstein-Rosen wave intensity $f_{(0)}$, with the
assumption that $V = o(1/r^2)$ as $r \rightarrow \infty$. We have neglected
boundary terms in this action, and will continue to do so for this paper,
because we are not interested in the dynamics corresponding to
evolution in the foliation label $t$, but
only in the constraints and the corresponding
internal hypertime dynamics, which are not affected by these terms.
The Hamiltonian form of the system is then given by
\begin{equation}{}\label{HamC}
S=\int dt\, \int dr\, \left( \pi_\gamma \dot{\gamma}
				+\pi_\phi \dot{\phi}
				+\sum_{i=0}^N \pi_{(i)} \dot{f}_{(i)}
				-\alpha {\cal H}
				-\beta {\cal H}' \right),
\end{equation}
where $\pi_\gamma$, $\pi_\phi$ and $\pi_{(i)}$ are the momenta
conjugate to $\gamma$, $\phi$ and $f_{(i)}$ respectively.
${\cal H}$ and ${\cal H}'$ are the superhamiltonian
\begin{equation}{}\label{superham}
{\cal H}= -4\pi_\gamma \pi_\phi
	+ \frac{1}{4}\left( \phi_{,rr} -\phi_{,r} \gamma_{,r} \right)
	+{\cal H}_f + \phi \exp (2\gamma) V
\end{equation}
and supermomentum
\begin{equation}{}\label{supermom}
{\cal H}' = \pi_\gamma \gamma_{,r} + \pi_\phi \phi_{,r}
	  - \pi_{\gamma,r} + {\cal H}'_f,
\end{equation}
and we have written ${\cal H}_f$ and ${\cal H}'_f$ for the $V=0$ part
of the superhamiltonian and supermomentum for the graviton and scalar
fields. These constraints do little for our intuition, so let us use
a variant of the Kucha\v{r} \cite {KK1} transformation on them by
defining the following embedding variables:
\begin{equation}{}\label{Tdef}
T^\pm (r) = 4 \int_r^\infty \pi_\gamma (r') dr' \pm \phi (r) +t.
\end{equation}
The conjugate momenta are then
\begin{equation}{}\label{Pdef}
\Pi_\pm (r) = \frac{1}{8} \frac{\partial}{\partial r}\left(
	\ln \left( \pm T^\pm_{,r} (r) \right)
	 - \gamma (r) \right)  \pm \frac{1}{2} \pi_\phi (r),
\end{equation}
which can be verified by an extensive calculation. We shall require
that $\pm T^\pm_{,r} >0$, which can be shown to correspond to the lack
of apparent horizons. It is then a simple matter to show
that, for $V=0$, $T^\pm$ are simply a set of null coordinates on the
classical solution spacetimes and approach the null coordinates
$t \pm r$ as $r \rightarrow \infty$ in all cases, when the appropriate
boundary conditions are satisfied \cite{KK1}.

Under the above transformation, the superhamiltonian becomes
\begin{eqnarray}{}\label{newHam}
{\cal H} & = & \Pi_+ (r) T^+_{,r} (r)
      - \Pi_- (r)  T^-_{,r} (r) + {\cal H}_f (r) \nonumber \\
& - &  \frac{1}{2} \left( T^+ (r) - T^- (r) \right) V (r)
	T^+_{,r} (r) T^-_{,r} (r)
	\exp \left( 8 \int_r^\infty dr'
	\left( \Pi_+ (r') + \Pi_- (r') \right) \right),
\end{eqnarray}
and the supermomentum becomes
\begin{equation}{}\label{newMom}
{\cal H}' = \Pi_+ (r) T^+_{,r} (r)
      + \Pi_- (r)  T^-_{,r} (r) + {\cal H}'_f (r),
\end{equation}
and we have used the boundary conditions to determine integration
constants in Equation~(\ref{newHam}). This is, at first sight,
a truly awful set of constraints, being highly non-linear and
non-local in the embedding momenta, and could not be used as the
basis for a hypertime quantum evolution equation, as they
would be infinite order in the hypertime! However, the
coupled integro-differential equations for $\Pi_\pm$ represented by
Equations~(\ref{newHam}) and~(\ref{newMom}) can be solved.
This surprisingly {\it unique} solution is arrived at by differentiating the
equations with respect to $r$, thus reducing the problem to a first
order ordinary differential equation, and then choosing integration
constants correctly to finally get the following:
\begin{equation}{}\label{Qform}
- \Pi_\pm = P_\pm \equiv \pm {\cal Q}\frac{T^\mp_{,r}}{8\left(
	T^+_{,r}-T^-_{,r} \right)}
	 \pm \frac{{\cal H}_f \pm {\cal H}'_f}{2 T^\pm_{,r}},
\end{equation}
where
\begin{equation}{}\label{Qdef}
{\cal Q}(r) = \frac{\partial}{\partial r} \left(
\ln \left( 1 +  \int_r^\infty V \frac{\partial}{\partial r'}
	\left( T^+ - T^- \right)^2 \exp \left(
	-4\int_{r'}^\infty h_f(r'') dr'' \right) dr' \right) \right),
\end{equation}
and
\begin{equation}{}\label{hfdef}
h_f = \frac{{\cal H}_f + {\cal H}'_f}{T^+_{,r}}
    - \frac{{\cal H}_f - {\cal H}'_f}{T^-_{,r}}.
\end{equation}

We have thus derived a hypertime formalism, corresponding to a
solution of the form given in Equation~(\ref{solnPi}), and have retrieved
an equation that represents first order propagation in the embedding
variables. It is important to note that this reduces to the results
by Kucha\v{r} \cite{KK3,KK1} when $V=0$ and the formalism becomes
local. The structure represented by Equations~(\ref{Qform})
to~(\ref{hfdef}) may look confusing, but it does have an understandable
form; The function ${\cal Q}$, although highly non-local, represents
the local interaction potential term in Equation~(\ref{superham}),
with the conformal factor $\gamma$ that results from the energy-momentum
density being generated by the scalar and graviton fields throughout
the spacetime. The formalism, as such, is therefore similar
to the Hamiltonian for
the reduced spherical models investigated by Unruh \cite{NOTES}
and Hajicek \cite{SSFII}. It should be noted, however, that
whereas the latter produce an evolution on a preferred single foliation,
the evolution represented by Equation~(\ref{Qform}), via
Equation~(\ref{HHam}), is for hypertime. This property therefore
allows us to analyze these models without making a specific
foliation choice. There is ongoing research into the
application of this basic technique to the entire family
of $2D$ gravity models.

The author would like to thank Redouane Fakir for his remarks on
this paper and Tamilyn Adams for encouragement.
This work is supported by the Natural Sciences and
Engineering Research Council of Canada.


\end{document}